\documentstyle[eqsecnum,aps,epsf]{revtex} 

\begin{document}

\twocolumn[\hsize\textwidth\columnwidth\hsize\csname@twocolumnfalse\endcsname

\title{Numerical study of the coupled time-dependent 
Gross-Pitaevskii equation: Application to Bose-Einstein
condensation}

\author{Sadhan K. Adhikari}
\address{Instituto de F\'{\i}sica Te\'orica, Universidade Estadual
Paulista, 01.405-900 S\~ao Paulo, S\~ao Paulo, Brazil\\}

\date{\today}
\maketitle
\begin{abstract}

We present a numerical study of the coupled time-dependent
Gross-Pitaevskii equation, which describes
the Bose-Einstein condensate of several types of trapped bosons at
ultralow temperature with both attractive and repulsive interatomic
interactions. The same approach is used to study both stationary and
time-evolution problems.  We consider up to four types of atoms in the
study of stationary problems. 
We consider  the time-evolution problems where the frequencies
of the traps or the atomic scattering lengths are suddenly changed in a
stable
preformed condensate. We also study the effect of periodically varying
these frequencies or scattering lengths on a preformed condensate.  These
changes introduce oscillations in the condensate which are studied in
detail. Good convergence is
obtained in all cases studied.

{\bf PACS Number(s): 02.70.Rw, 02.60.Lj, 03.75.Fi}

\end{abstract}

\vskip1.5pc]
 \newpage \section{Introduction}

The experimental detection \cite{1} of Bose-Einstein condensation (BEC) 
at ultralow temperature in dilute bosonic atoms (alkali and hydrogen
atoms)  employing magnetic traps  have spurred intense theoretical
activities  on various aspects of the condensate
\cite{3,5,6,7,cst,11,13}. Many properties of the condensate are
usually described by the nonlinear mean-field Gross-Pitaevskii (GP)
equation \cite{11,8}.  The GP equation in both time-dependent and
independent
forms is formally similar to the Schr\"odinger equation including  a
nonlinear
term \cite{5,6,cst}.

More recently, there has been experimental realization of BEC involving
two types of atoms \cite{excpl1,matt,excpl2}. In the actual
experiment
$^{87}$Rb
atoms
formed in the $F=1$, $m=-1$ and $F=2$, $m=1$ states by the use of a laser
served as two different species of atoms, where $F$ and $m$ are the total
angular momentum and its projection \cite{matt}. In another experiment
a coupled BEC was formed with the $^{87}$Rb
atoms in the $F=1$, $m=-1$ and $F=2$, $m=2$ states.
 \cite{11,excpl2}. Experimentally, it is possible to use the same magnetic
trap to confine the atoms in two quantum states, which makes this study
easier technically compared to the formation of a BEC with two different
types of atoms requiring two different trapping mechanisms.  
It has also been found in these studies \cite{matt,excpl2} that the
$^{87}$Rb
atoms have a
repulsive interaction in all three states considered above.  These
experiments initiated theoretical activities in multicomponent BEC
described by the coupled GP equation
\cite{cpl1}.

A numerical study of the time-dependent coupled GP equation is interesting
as this can provide solution to many
time-evolution problems involving more than one types of atoms forming a
BEC. The solution of the coupled nonlinear GP equation is
nontrivial \cite{ns} and here  we undertake the challenging task of the
numerical study of these time-evolution problems.

In a multi-component BEC the main feature  is the coupling between
different types of condensates, which can lead to new effects associated
with this novel and surely much richer situation inexistent in a
single-component BEC. We list  some interesting   possibilities below
which we shall investigate numerically in this study.  It is
possible to have a distinct trapping frequency for each of the components,
one of which can be suddenly altered experimentally.  Of course, this
change would affect the component of the BEC directly trapped by this
field.  However, it is more interesting to study how this sudden change
affects the other component of BEC not directly trapped by this field.
Also it is possible to suddenly vary \cite{cor} the atomic scattering
length of one of the species and study its effect on the other component.
The above variation of one of the trap frequencies or scattering
lengths can be carried through in a periodic fashion and its effect on the
other component can be studied. The studies mentioned above are peculiar
to a coupled BEC and are of interest as it is now possible
to vary both the trap frequencies \cite{td} as well as the scattering
lengths \cite{11,matt,cor} both abruptly or in a periodic fashion.
These effects do not have any analogue in the uncoupled BEC and we
motivate the present investigation with special emphasis on these
effects in this study. 
We investigate  these time-evolution
problems using a set of two coupled GP equations in the purely repulsive
case. 
In addition we study the stationary solution to
the coupled GP equation describing a 
multi-component  condensate, where we consider up to four components.

We solve the coupled BEC problem using the time-dependent coupled GP
equation in cases of attractive and repulsive atomic
interactions by
discretization
 with the Crank-Nicholson-type rule
complimented by the known boundary conditions at origin and
infinity \cite{koo}. 
This procedure leads to good convergence for both the stationary and
time-evolution 
problems.

First we consider stationary coupled
condensates under the action of trap potentials.  Stable and converged
numerical results are obtained for up to four coupled equations in the
repulsive case and two  in the attractive case. 
The time-dependent GP
equation is directly solved to obtain the full time-dependent solution
in the case of  stationary problems, from which a trivial time
dependent phase factor  is separated and the stationary
solution obtained as in the uncoupled case \cite{7}.

We also study the numerical
stability of the calculational scheme which is more difficult to obtain
when the nonlinearity is large.  
For this purpose we only consider
repulsive interatomic interactions in the time-evolution problems where
condensates with large 
nonlinearity can be formed. In the case of attractive interaction 
a large nonlinearity leads to the collapse of the condensate \cite{13}.

In Sec. II we present  the  coupled
time-dependent  GP equation which we use. In Sec. III we
describe the numerical method in some
detail. In Sec. IV we report the numerical results for the stationary
case and in Sec. V we report a study of three types of time-evolution
problems.
Finally, in Sec. V
we give a summary of our investigation.

\section{Nonlinear Coupled Gross-Pitaevskii Equation}

The GP  equation \cite{8} for a
coupled trapped Bose-Einstein
condensate  at zero temperature is written as
\cite{cpl1}
\begin{eqnarray}\label{aa} \biggr[ -\frac{\hbar^2}{2m}\nabla^2
&+&\frac{1}{2}c_im\omega^2 r^2 +\sum_{l=1}^M g_{il}N_l|\Psi_l({\bf
r},\tau)|^2\nonumber \\
&-&i\hbar\frac{\partial
}{\partial \tau} \biggr]\Psi_i({\bf r},\tau)=0,   \end{eqnarray} 
where  $\Psi_i({\bf r},\tau)$ at position ${\bf r}$ and time $\tau $ 
is the wave function for the component $i$ of the condensate,
$m$
is
the mass of a single bosonic atom, $N_l$ the number of condensed atoms of
type  $l$,  $M$ the number of types of atoms, $c_i m\omega^2 r^2/2$ the
attractive harmonic-oscillator trap
potential, $\omega$ the oscillator frequency.  The
parameter $c_i$ has been introduced to 
independently modify the frequency of the harmonic oscillator trap for
each type of atoms. 
Here $g_{il}=4\pi\hbar^2a_{il}/m$ is the coupling constant for elastic
interaction between atoms of types $i$ and $l$, where $a_{il}$ is the
corresponding scattering length.
The masses of different types of atoms are taken to be
equal,  
as this is necessary while considering 
the coupled BEC formed of different spin states of $^{87}$Rb $-$ one of
the most important realization to date. 
In this
work we
shall not allow the transition of one type of atoms  of the BEC to the
other
and take the number of atoms of each component to be constant as in the
experiment of Ref. \cite{excpl2}.

Here we shall be interested in the spherically symmetric solution
$\Psi_i({\bf r},\tau)\equiv \psi_i({ r},\tau)$
to Eq.  (\ref{aa}),
 which can be written  as \begin{eqnarray}\label{cc} \biggr[
-\frac{\hbar^2}{2m}\frac{1}{r}\frac{\partial^2 }{\partial
r^2}r &+& \frac{1}{2}c_im\omega^2 r^2 +\sum_{l=1}^M
g_{il}N_l|\psi_l({
r},\tau)|^2 \nonumber 
\\
&-& i\hbar\frac{\partial}{\partial \tau}\biggr]
\psi_i(r,\tau)=0,
\end{eqnarray}
The above limitation to the spherically symmetric solution
(in the zero angular momentum state) reduces the coupled GP equation 
 to a one-dimensional coupled partial differential equation.

As in Refs. \cite{6,7} it is convenient to use dimensionless variables
defined by $x = \sqrt 2 r/a_{\mbox{ho}}$ , and $t=\tau \omega, $
where
$a_{\mbox{ho}}\equiv \sqrt {\hbar/(m\omega)}$, and $
\phi_i(x,t) = x\psi_i(r,\tau ) (\sqrt 2\pi a_{\mbox{ho}}^3)^{1/2}$ . In
terms of these
 variables Eq. (\ref{cc})  becomes
\begin{eqnarray}\label{e}\biggr[ -\frac{\partial^2 }{\partial
x^2} &+& \frac{1}{4}c_ix^2 +\sum_{l=1}^M n_{il}
\frac{|\phi_l({x},t)|^2}{x^2}\nonumber \\
& - & i\frac{\partial
}{\partial t} \biggr]\phi_i({ x},t)=0, \end{eqnarray}
where
$n_{il}\equiv 2\sqrt 2 N_l a_{il}/a_{\mbox{ho}}$ is the reduced number of
particles and this number could be negative (positive) when the
corresponding
scattering length is negative (positive) representing an attractive
(repulsive) interatomic interaction.
The normalization  of the wave
function is $ \int_0 ^\infty |\phi_i(x,t)|
^2 dx
 = 1  $ 
and  its
root-mean-square (rms)
radius $x^{(i)}_{\mbox{rms}}$   is given  by \begin{equation}\label{7}
x^{(i)}_{\mbox{rms}}=
\left[\int_0
^\infty x^2 |\phi_i(x,t)| ^2 dx\right]^{1/2}.  \end{equation}

\section{Numerical Method}

To solve Eq. (\ref{e}) numerically one needs the proper boundary
conditions at $x\to 0$ and $\infty$. For a confined condensate, for a
sufficiently large $x$, $\phi_i(x,t)$ must vanish asymptotically. Hence
the nonlinear term proportional to $|\phi_i(x,t)|^3$ can eventually be
neglected in the GP equation for large $x$. Consequently the the
asymptotic form of the physically acceptable solution is given by $
\lim_{x \to \infty} |\phi_i(x,t)|\sim \exp (-{x^2}/{4}).$ Next we consider
Eq. (\ref{e}) as $x\to 0$. The nonlinear term approaches a constant in
this limit because of the regularity of the wave function at $x=0$. Then
one has the condition $ |\phi_i(0,t)|=0.$

A convenient way to solve Eq. (\ref{e}) numerically is to discretize it in
both space and time and reduce it to a set of algebraic equations which
could then be solved by using the known asymptotic boundary conditions.
The procedure is similar to that in the uncoupled case \cite{7}.  We
discretize Eq. (\ref{e}) by using a space step $h$ and time step $\Delta $
with a finite difference scheme using the unknown ${(\phi_i)}^k_p$ which
will be approximation of the exact solution $\phi_i(x_p,t_k)$ where $x_p=
p h$ and $t_k=k\Delta $.  The time derivative in Eq. (\ref{e}) involves
the wave function at times $t_k$ and $t_{k+1}$. As in the uncoupled case
we express the wave functions and their derivatives in Eq. (\ref{e}) in
terms of the average over times $t_k$ and $t_{k+1}$ \cite{koo} and the
resultant scheme leads to accurate results and good convergence. In
practice we use the following 
Crank-Nicholson-type  scheme \cite{koo}  to
discretize the
partial differential equation (\ref{e}) 
 \begin{eqnarray} i\frac{(\phi_i)_p^{k+1}-(\phi_i)_p^{k}
}{\Delta
} &=& -\frac{1}{2h ^2}\biggr[(\phi_i)^{k+1}_{p+1}-2
(\phi_i)^{k+1}_{p}+(\phi_i)^{k+1}_{p-1} \nonumber \\ &+&
(\phi_i)^{k}_{p+1}-2
(\phi_i)^{k}_{p}+(\phi_i)^{k}_{p-1}\biggr]\nonumber \\ 
&+&\frac{1}{2}\left[\frac{c_ix_p^2}{4}+\sum_{l=1}^M
n_{il}\frac{|(\phi_l)_p^{k}|^2}{x_p^2}
\right]\nonumber \\
&\times&
[(\phi_i)_p^{k+1}+(\phi_i)_p^k]. \label{f} \end{eqnarray}

Considering that
the wave function components $\phi_i$  are  known at time $t_k$, Eq.
(\ref{f}) is an equation in
the unknowns $-$ $(\phi_i)_{p+1}^{k+1},(\phi_i)_p^{k+1}$ and
$(\phi_i)_{p-1}^{k+1}$. 
In a lattice of $N$ points Eq. (\ref{f}) represents a tridiagonal set for
$p=2,3,...,(N-1)$ for a specific component $\phi_i$. This set has a unique
solution if the wave functions at
the two end points $(\phi_i)_{1}^{k+1}$ and $(\phi_i)_{N}^{k+1}$ are
known \cite{koo}. In the
present problem these values at the end points are provided by the known
asymptotic conditions.

To find the ground state of the condensate we start with the analytically
known properly normalized wave functions of the uncoupled harmonic
oscillator problems described by Eq. (\ref{e})  with $n_{il} =0$.  We then
repeatedly propagate these solutions in time using the
Crank-Nicholson-type algorithm (\ref{f}). Starting with $n_{il}=0$, at
each time step we increase or decrease the nonlinear parameter $n_{il}$ by
an amount $\Delta_1$.  This procedure is continued until the desired final
value of $n_{il}$ is reached.  The resulting solution is the ground state
of the condensate corresponding to the specific nonlinearity.

The time-dependent approach is the most suitable for solving
time-evolution problems. In the present study we only consider evolutions
problem starting from a stable condensate at $t=0$. In these cases the
stationary problem is solved first and the wave function so obtained
serves as the starting wave function for the time-evolution problem.

\section{Results for the Stationary Problem}

First we consider the stationary ground-state solution of Eq. (\ref{e}) 
in cases of both attractive and repulsive interactions using two and four
coupled equations.  The numerical integration was performed up to
$x_{\mbox{max}}=15$ with $h=0.0001$ using time step $\Delta=0.05$ and the
parameter $\Delta_1 = 0.01$.  After some experimentation we find that good
convergence is obtained with parameters $\Delta$ and $\Delta_1$ near these
values.  The convergence is
fast for small nonlinearity. The final convergence of the scheme breaks
down if nonlinearity is too large. In practice these difficulties start
for $n_{ii}>20$ for the ground state for a repulsive interaction in a
computational analysis in double precision.  For an attractive interaction
the coupled GP equation does not sustain a large
nonlinearity $|n_{il}|$  and leads to collapse. Except for values of
nonlinearity near collapse, the GP equation in the attractive case leads
to good convergence.

\subsection{Repulsive Atomic Interaction}

In most of the experimental realization of BEC in trapped atoms, the
interatomic interaction is repulsive and we consider this case first.
We consider the simple case of two coupled GP
equations in the case of repulsive  interaction with (a)
$n_{11}=n_{22}=10$,
$n_{12}=n_{21}=5$,
$c_1 =1$ and  $c_2=0.25$, 
(b)
$n_{11}=n_{22}=5$,
$n_{12}=n_{21}=100$,
$c_1 =1$ and  $c_2=0.25$, and (c)
$n_{11}=n_{22}=10$,
$n_{12}=n_{21}=5$, and 
$c_1 =c_2=1$. In this case all interactions are repulsive corresponding to
the positive sign of all $n_{il}\equiv 2\sqrt 2
N_{i}a_{il}/a_{\mbox{ho}}$. Although these parameters are in dimensionless
units, it is easy to associate them to an actual physical problem  of
experimental interest. For
example, for the mixture of $|F=1, m=-1\rangle$ and 
$|F=2, m=1\rangle$  $^{87}$Rb states, the ratio of scattering lengths 
$a_{|1,-1\rangle}/a_{|2,1\rangle}=1.062$ \cite{matt}. If we label  the 
${|1,-1\rangle}$ state by 1 and the ${|2,1\rangle}$ state by 2, and
consider 
$a_{11}/a_{\mbox{ho}}\simeq a_{22}/a_{\mbox{ho}}\simeq 0.002$, then 
$n_{11}=n_{22}=10$ corresponds to  $N_1\simeq
N_2\simeq 1770$. This estimate gives an idea of the actual experimental 
condition that the present set of parameters simulate. The three models
considered above can simulate actual experimental situations composed of
two states of $^{87}$Rb.
 The
different values
of $n_{12}$ and $n_{21}$ considered above corresponds to different
possible unknown repulsive interactions among the two species of
condensates.
It is realized from the coupled GP equation that in
case of model (c) $\phi_1 =\phi_2$.  
We show results for the two components of the wave function for
two
sets of values of $\Delta_1$: 0.01 (full line) and 0.015 (dashed line) in
Fig. 1. The difference
between the two sets of results increases as the nonlinearity of the
GP equation given  by $n_{il}$ increases, e.g., for the case (b) above
compared to (a). The difference 
reduces to zero as the nonlinearity
decreases.

Next we consider the more complicated case of four coupled GP equations
with repulsive atomic interaction.
This is a purely theoretical case with no experimental analogue, as all
experiments to date are limited to two coupled condensates only. 
In
this case the numerical method works in the same fashion and good
convergence is obtained with moderate values of nonlinearity. 
Again we consider the case of repulsive
interaction between all possible pairs. In this case in the four-component
model we take $n_{11} = 4$, $n_{22} = 5$, $n_{33} = 6$, $n_{44}
= 8$, and $n_{il} = 2, i\ne l$; $c_1= 4$, $c_2= 1$, $c_3= 4$, and $c_4=
0.25$. The solution for the wave-function components obtained with $\Delta
= 0.05$ and $\Delta_1= 0.01$ (full line), and 
 $\Delta_1= 0.015$ (dashed line) 
are shown in Fig. 2. The maximum difference between the two 
calculations is about 6$\%$ for the largest component ($\phi _ 1$) near
$x=0$, although the average difference is much smaller as can be seen in
Fig. 2.

\subsection{Attractive Atomic Interaction}

The case of attractive atomic interaction demands special attention as 
one can have the phenomenon of collapse in this case. 
We consider the case of two coupled GP equations with attractive
interactions between like atoms $ii$, $i=1,2, $ and with repulsive
interactions between unlike atoms $ij$, $i\ne j.$ In this case some of
the atomic interactions are attractive in nature. 
Consequently,  with large (attractive)
nonlinearity the system may undergo collapse and for stable stationary
solution of the GP equation the nonlinearity should be maintained small.
We consider the following three cases: (a) $n_{11}=-1$, $n_{22}= -1.5$,
$n_{12}=n_{21}=4$, $c_1=4$, and $c_2=1$;  (b) $n_{11}=-1$, $n_{22}= -1.5$,
$n_{12}=n_{21}=80$, $c_1=4$, and $c_2=1$;  (c) $n_{11}=-1$, $n_{22}=
-1.5$, $n_{12}=n_{21}=100$, $c_1=4$, and $c_2=1$.  The only observed case
of BEC with attractive  interaction is the case of $^7$Li with
$a/a_{\mbox{ho}} \simeq -0.0005$ \cite{11}. If we label this state by 1
then $n_{11}=-1$ corresponds to the actual particle number $N_1=700$. In
an
uncoupled condensate of $^7$Li the BEC collapses for more than 1400 atoms.
Although there has been no experimental realization of coupled BEC in the
case of attractive interaction, the parameters cited above may simulate
the  BEC of ground-state atoms of $^7$Li coupled to one of its excited
states, where the atomic interaction is also attractive. If we assume that
the excited-state atoms have the same value of $a/a_{\mbox{ho}}$ as in the
ground state then $n_{22}=-1.5$ corresponds to the number of atoms
$N_2=1500$ in the excited state where the excited state is labeled by the
index 2.   

 The wave-function
components in this case are shown in Fig. 3.  As for stable stationary
solution the nonlinearity in this case has to be smaller than in the
purely repulsive case, numerically it is easier to obtain precise
solution except for values of the nonlinearity close to (and beyond)
collapse. 
The parameters above in cases (a), (b), and (c) are chosen to illustrate
the collapse of the system arising from the divergence of the first
component of the wave function ($\phi _1$). The nonlinear couplings
$n_{12}$ and $n_{21}$ are increased as we move from case (a) to (b) and
then to (c). Other parameters of the model are kept fixed.  Because of the
attractive interactions in this case, the system moves towards collapse as
we move from case (a) to (c) through (b).  The central density
corresponding to $\phi_1$ increases (eventually tends to infinity) and the
rms radius of the system decreases (eventually tends to zero) with the
increase of nonlinearity. This is clear from a comparison of Fig. 3 with
Figs. 1 and 2. The very large central value of the wave function $\phi_1$
($\sim 35$) and its small radial extension indicates a large central
density and a small rms radius.

Finally, we consider the case of two interacting systems with all
interactions repulsive. In this case the system is more vulnerable to
collapse if the nonlinearity is large. We consider the following three
sets of parameters in this case for which we show the solution in Fig. 4: 
(a) $n_{11}=-1$, $n_{22}= -1$, $n_{12}=n_{21}=-0.4$, $c_1=4$, and
$c_2=0.25$; (b)  $n_{11}=-1$, $n_{22}= -1$, $n_{12}=n_{21}=-0.5$,
$c_1=4$, and $c_2=0.25$; (c) $n_{11}=-1$, $n_{22}= -1$,
$n_{12}=n_{21}=-0.55$, $c_1=4$, and $c_2=0.25$. These parameters simulate 
the possible coupled BEC composed of the attractive ground and excited
states of $^7$Li where the interaction between a ground- and
an excited-state atom is also taken to be attractive. It is possible to 
calculate the number of the two types of atoms as in the discussion
related to Fig. 3.
Here the
nonlinearity increases as we move from case (a) to (b) and then to (c),
and consequently, the wave-function components become more and more
localized with a large central density and small rms radii signaling the
onset of 
collapse of the system. This is clear from Fig. 4. In case (c) the
nonlinearity is the highest and one is closer to collapse. However, there
is a
difference between the two collapses shown in Figs. 3 and 4. In Fig. 3 the
route to collapse is manifested through a singular behavior of component
$\phi_1$ of the wave function; in Fig. 4 both  components $\phi_1$ and
$\phi_2$ exhibit the singular behavior. The spacial extension of the wave
function
components close to collapse in Figs. 3 and 4 is much smaller than the
wave function components in the purely repulsive cases shown in Figs. 1
and
2.

\subsection{Estimate of Numerical Error}

It is appropriate to comment quantitatively on the numerical accuracy of
the present method.  If we iterate the final solution in time without
changing the nonlinearity, the numerical result keeps on oscillating with
a small amplitude around the converged value. This oscillation gives a
good estimate of the numerical error of the method. This error manifests
in a different way in Fig. 1, where we have varied $\Delta_1$.  The
numerical solution of the time-dependent method is independent of the
space step $h$ provided that a typical value around $h=0.0001$ is employed
as in the present study. No visible difference in the solution is found if
$h$ is increased by a factor of two. However, the above oscillations with
time iteration are sensitive to the parameters $\Delta$ and $\Delta_1$. 
The values of these parameters ($\Delta =0.05$ and $\Delta_1=0.01$)  are
chosen to minimize the oscillation of the results with time iterations. 
The oscillation increases if larger or smaller values of one or both of
these parameters are employed and can really be large if an improper value
of step $\Delta$ or $\Delta_1$ is chosen. This oscillation is quite
similar to that in the uncoupled case studied in detail in Ref. \cite{7}.

From Figs. 1 and 2 we find that
the
error in the wave function component $|\phi_i (x,t)/x|$ as a function of
$x$ is the
largest at $x=0$.  In Table I we show the percentage error of $\lim_{x\to
0}|\phi_i (x,0)/x|$, defined by ${\cal E} \equiv
100[|\phi_i(0,t)|-|\phi_i(0,0)|]
/|\phi_i(0,0)|$ at those iterations where this error is maximum. For
illustration we consider the models (a) discussed in Figs. 1 and 4. At
positive ${\cal E}$ there is overshooting and at negative ${\cal E}$ there
is undershooting. Between a positive ${\cal E}$ and a negative ${\cal E}$
there is a zero of ${\cal E}$ denoting zero error.   We find that the
wave functions oscillate with time around the stationary solution. The
maximum reported error is about 6$\%$. Considering that we are dealing
with coupled nonlinear equations these errors are well within the
acceptable
limit.  The errors shown in Table I would also be the typical errors in
time-evolution problems with same nonlinearity which we study in Sec. V.
From Table I we see that the period of oscillation of the
result varies from one case to another.  However, the error increases as
the nonlinearity increases  or as the system approaches collapse in the
case of attractive interaction. For example, in models (c) of Figs. 3 and
4, which are close to collapse corresponding to almost maximum permissible
nonlinearity, the error increases quickly with time iteration  and a
large
numerical error could be  generated easily.

\section{Results for the Time-Evolution Problem}

Now we consider three types of  time-evolution problems,
 some of which could possibly be studied experimentally.
As the
repulsive case leads to more stable configuration
of the condensate, in this study we consider only this case in the process
of time
evolution. The attractive case of coupled BEC is also very interesting
from a physical point of view because of the occurrence of collapse.
We have performed a study of the dynamics of collapse in coupled BEC using
the
present numerical 
method, which will be reported elsewhere. 
The two types of parameters that can be varied in the time-evolution
study are the 
frequencies of the harmonic oscillator traps and the different scattering
lengths. 
Recently, it has been possible to vary the scattering length
experimentally by
varying an external field \cite{matt,excpl2}. It is also possible to vary
the trap frequency by varying the  currents in the magnets
responsible for confinement \cite{td}.

In the first type of problems we consider a sudden change of the harmonic
oscillator frequencies or scattering lengths at  $t=0$ and study its
effect on a preformed
condensate. 
In the second type we study the effect of  a periodic
temporal variation of these
frequencies on a preformed condensate. Finally, we study the effect of a
periodic
temporal variation of the scattering lengths on the preformed condensate.
In all cases we take the preformed condensate as the one described by the
model (a) of Fig. 1. 
We have commented before that the parameters of this model 
can simulate the coupled BEC composed of the ground and an excited state
of $^{87}$Rb, which gives a motivation for this choice.
When we
implement these time-dependent perturbations, the system starts to
oscillate (grow and shrink) with time.  The corresponding evolution  can
be studied
best through  the rms radii  \cite{11} which execute
periodic oscillation with time. 

\subsection{Sudden Change of Trap Frequency or Scattering Length}

By varying the external fields it is possible to vary the harmonic
oscillator trap frequency of the
confining trap as well as the atomic interactions (scattering lengths)
\cite{cor}. 
First we consider a sudden change of both the trap frequencies  on the
preformed coupled stationary BEC state corresponding to model
(a) of Fig. 1. 
We set the reduced time $T\equiv t/0.05=0$ when we start
the time evolution. As the time step $\Delta$ is 0.05, $T$ is just the
number of iterations. In this model we have two different trap frequencies
for the two components given by $c_1=1$ and $c_2=0.25$. 
In the first case at $T=0$ we suddenly interchange
the constants $c_1$ and $c_2$, i.e., we set $c_1= 0.25$ and $c_2=1$ and
study the time evolution. The evolution of the rms radii corresponding to
$\phi_1$ and $\phi_2$ are shown in Fig. 5 (a). Both rms radii execute
oscillations. However, that corresponding to $\phi_1$  has a
much larger amplitude. The periods of oscillation of the two radii are
different.

Next we consider a sudden change in one of  the trap frequencies 
on the preformed condensate at $T=0$ corresponding to
model (a) of Fig. 1 with $c_1=1, c_2=0.25$. At $T=0$,  we set $c_1=c_2=1$,
which
corresponds to
parameters of
model (c) of Fig. 1. The evolution of the rms radii is shown in Fig. 5
(b). Although in the stationary configuration of model (c) of Fig. 1,
$\phi_1=\phi_2$, this condition is
never attained in this evolution problem. The system keeps on oscillating
indefinitely with time. The oscillation shown in Figs. 5 (a) and (b) has
nothing to do with the
nonlinear or coupled nature  of the problem. Similar oscillation also
appears in an uncoupled linear oscillator when the trap frequency is
suddenly changed. In the present coupled nonlinear problem  both rms radii
execute
oscillations with time.  However, when the amplitude of oscillation of
one of the components increases, that of the other decreases. This
behavior denotes the transfer of kinetic energy from one component to the
other. 

Now we study the effect of a sudden change of the scattering length(s) on
the preformed condensate \cite{cor}. We  
consider the problem when the parameters of model (a) of Fig. 1 are
suddenly changed to those of model (b) of Fig.  1 at $T=0$. This is
achieved by changing the nonlinearities  suddenly at $T=0$ from
$n_{11}=n_{22}=10, n_{12}=n_{21}=5$ to $n_{11}=n_{22}=5,
n_{12}=n_{21}=100$ with a
variation of the external field which controls the scattering length(s).
In this
case the oscillations of the system are shown in Fig. 5
(c) where we plot 
the time evolution of the rms radii of the two components. 
Both
components of the condensate execute oscillations but with different
frequencies and amplitudes. One of the components execute giant
oscillations with large amplitude, whereas the amplitude of the other is
much smaller.

Finally, we consider the case when one of the
trapping potentials is switched off at $T=0$ on the preformed condensate
of Fig. 1 model (a) by setting $c_2=0$. 
The oscillation in this case is
shown in Fig. 5 (d) where we plot the two rms radii. In the absence of the
trapping potential the second
component of the condensate can no longer remain localized in space.
However, it does not expand monotonically before evaporating.  It
starts to execute giant oscillation  and eventually
escapes to
infinity. 
Similar oscillation was found in the case of an uncoupled 
BEC when the trapping potential was removed \cite{7}. 
The first component essentially remains unchanged during the
process under the action of the unchanged trap potential. The minor
oscillation of the rms radii of the first component is due to  the
coupling to the expanding second component.

\subsection{Periodic Oscillation of Trap Frequency}

Instead of making a sudden change in the parameters of the model, next we
introduce periodic oscillation in some of the parameters of the model for
$T\ge 0$ and study the consequence on the system. We introduce
a periodic
variation in the parameters $c_i$ which are related to the harmonic
oscillator trap frequencies. Experimentally, this
variation is possible via a variation of the external fields which are
controlled by currents.  

We again 
consider at $T=0$ the preformed condensate of the model (a) of Fig.  1.
First we consider the variation $c_1= 1-0.5\sin(\pi T/20)$ which
corresponds to varying the frequency of the first trap. The resultant
variation of the two rms radii are shown in Fig. 6. The first radius
(full line)  oscillates more rapidly with larger amplitude and frequency
than the second radius (dashed line). This is reasonable as we are
directly varying the first frequency in this case. The rms radius of the
second wave function feels the effect through its coupling to the first
component. We also varied both the papameters $c_1$ and $c_2$ in a
periodic fashion which corresponds to varying both the frequencies. In
this case both the rms radii execute oscillation. However, no interesting
effect is observed and we do not show the details of that oscillation
here.

\subsection{Periodic Oscillation of Scattering Length}

Now we study the effect of a periodic variation of the scattering
length(s) of the system on a preformed condensate.  In our formulation
this corresponds to a periodic variation of the parameters $n_{il}$. This
variation of the atomic interactions or the scattering lengths is now
feasible experimentally \cite{cor}.  We consider the periodic
variation in one of the scattering lengths ($a_{11}$) by setting $n_{11}=
1-0.5\sin(\pi T/20)$ for $T\ge 0$ on the preformed condensate of model (a)
of Fig. 1. The resultant oscillation of the rms radii are shown in Fig. 7. 
This variation corresponds to a variation of the atomic interaction
among atomic states of the first type. Consequently, the rms radii of the
first component of the BEC executes pronounced oscillation with moderate
amplitude. There is no direct variation in the parameters of the second
component.  The second component of the condensate feels the effect of
variation of $n_{11}$ through the coupling to the first component. Because
of this secondary effect the second component also executes oscillation as
can be seen from its rms radii in Fig. 7, albeit with a much smaller
amplitude compared to the first component.

We also  considered a periodic variation of the scattering length between
one atom of each type ($a_{12}=a_{21}$)
by setting
$n_{12}= n_{21}=
0.5-0.25\sin(\pi T/20)$ on the same preformed condensate for  $T \ge 0$
and studied the 
resultant oscillation of the rms radii. However, no interesting behavior
was observed and we do not show details of that oscillation.

\section{Summary}

In this paper we present a numerical study of the coupled
time-dependent Gross-Pitaevskii
equation for BEC in three space dimensions under the action of  harmonic
oscillator trap potentials  with attractive and repulsive
interparticle interactions between different types of atoms
\cite{cpl1}.

The time-dependent coupled GP equation is solved by discretizing it using
a Crank-Nicholson-type scheme \cite{7,koo}. This method leads to good
convergence for small nonlinearity. 
Numerical difficulty appears
for large
nonlinearity ($n_{il}>20$). For medium nonlinearity, the accuracy of the
method can be increased by reducing the space step $h$.

The ground-state stationary wave functions are found to be sharply peaked
near the
origin for attractive interatomic interaction for larger nonlinearity 
(Fig. 4). For a repulsive
interatomic
interaction the wave function extends over a larger region of space
(Figs. 1 and 2). In
the case of an attractive potential, the rms radii decrease with an
increase of nonlinearity. There could be a collapse for attractive
interaction when the nonlinear parameters $n_{il}$ are increased as in
the uncoupled case \cite{13}. In the purely repulsive case we solved two
and four coupled GP equations. In problems involving attraction we solved
only the two coupled GP equations.

In addition to the stationary problem we studied three types of evolution
problems. A stable coupled condensate is considered at $T=0$ on which a
time-dependent perturbation is introduced. Two types of perturbations were
considered on a two-component condensate with purely repulsive
interactions. In the first type a sudden change in the parameters
related to the frequencies of the trap and the
scattering lengths
was introduced.
 In the second
type a periodic variation of the different scattering lengths and the 
frequencies of the harmonic oscillator trap was introduced. In all cases 
the condensates execute periodic oscillation which is studied via the time
evolution of the rms radii as in the uncoupled case \cite{11}. 
We conclude that the present time-dependent approach is very suitable for
studying both the stationary and time-evolution problems of a coupled
BEC.

The work is supported in part by the CNPq and FAPESP of Brazil.

\newpage

{\bf Figure Caption:}

1. Wave function components  $\phi_1$ (label 1) and $\phi_2$ (label 2)
for
two coupled GP equations with (a)
$n_{11}=n_{22}=10, $ $n_{12}=n_{21}=5, $
$c_1=1$, $c_2=0.25$;  (b) 
$n_{11}=n_{22}=5, $ $n_{12}=n_{21}=100,
$
$c_1=1$, $c_2=0.25$; and (c) $n_{11}=n_{22}=10, $ $n_{12}=n_{21}=5,
$
$c_1=1$, $c_2=1$
calculated with  $\Delta = 0.05$ and $\Delta_1=0.01$
(full line); $\Delta = 0.05$ and  $\Delta_1=0.015$ (dashed line). In case
(c) only the results
for $\Delta_1=0.01$ are shown.

2.  Wave function components  $\phi_1$ (label 1), $\phi_2$ (label 2), 
$\phi_3$ (label 3), and $\phi_4$ (label 4)
for
four coupled GP equations with  $n_{11} = 4$, $n_{22} = 5$, $n_{33} = 6$,
$n_{44}
= 8$, and $n_{il} = 2, i\ne l$; $c_1= 4$, $c_2= 1$, $c_3= 4$, and $c_4=
0.25$ calculated with $\Delta =0.05$ and $\Delta_1=0.01$ (full line);
$\Delta =0.05$ and
$\Delta_1=0.015$ (dashed line). 

3. Wave function components  $\phi_1$ (label 1) and $\phi_2$ (label 2)
for
two coupled GP equations with 
(a) $n_{11}=-1$, $n_{22}= -1.5$, $n_{12}=n_{21}=4$, $c_1=4$, and
$c_2=1$ (dashed-dotted line);
(b) $n_{11}=-1$, $n_{22}= -1.5$, $n_{12}=n_{21}=80$, $c_1=4$, and
$c_2=1$ (dashed line);
(c) $n_{11}=-1$, $n_{22}= -1.5$, $n_{12}=n_{21}=100$, $c_1=4$ , and
$c_2=1$ (full line) calculated with $\Delta =0.05$ and $\Delta_1=0.01$.

4. Wave function components  $\phi_1$ (label 1) and $\phi_2$ (label 2)
for
two coupled GP equations with 
(a) $n_{11}=-1$, $n_{22}= -1$, $n_{12}=n_{21}=-0.4$, $c_1=4$, and
$c_2=0.25$ (dashed-dotted line); (b)  $n_{11}=-1$, $n_{22}= -1$,
$n_{12}=n_{21}=-0.5$,
$c_1=4$, and $c_2=0.25$ (dashed line); (c) $n_{11}=-1$, $n_{22}= -1$,
$n_{12}=n_{21}=-0.55$, $c_1=4$, and $c_2=0.25$ (full line) 
calculated with $\Delta =0.05$ and $\Delta_1=0.01$.

5. The rms radii of the two components $\phi_1$ (full line) and $\phi_2$
(dashed line) of the wave function at different reduced times
$T\equiv t/0.05$ for the oscillating condensate when  on the preformed
condensate of model (a) of Fig. 1 we suddenly inflict the following
changes: (a) $c_1= 0.25$ and $c_2=1$; (b) $c_1=c_2=1$; (c)
$n_{11}=n_{22}=5$; $n_{21}=n_{12}=100$; (d) $c_2 =0$ calculated with
$\Delta =0.05$
and $\Delta_1=0.01$. All other parameters are maintained unchanged.

6. The rms radii of the two components $\phi_1$ (full line) and $\phi_2$
(dashed line) of the wave function at different reduced times $T\equiv
t/0.05$ for the oscillating condensate when on the preformed condensate of
model (a) of Fig. 1 we suddenly inflict the following change: $c_1 =
1-0.5\sin (\pi T/20)$ calculated with $\Delta =0.05$ and
$\Delta_1=0.01$. All other parameters are maintained unchanged. 

7. The rms radii of the two components $\phi_1$ (full line) and $\phi_2$
(dashed line) of the wave function at different reduced times $T\equiv
t/0.05$ for the oscillating condensate when on the preformed condensate of
model (a) of Fig. 1 we suddenly inflict the following change:  $n_{11}
=
1-0.5\sin (\pi T/20)$
 calculated with $\Delta =0.05$ and
$\Delta_1=0.01$. All other parameters are maintained unchanged.

\vskip .2cm
\newpage
\onecolumn

{Table I: Percentage error  ${\cal E}\equiv 100
[|\phi_i(0,T)|-|\phi_i(0,0)|]/|\phi_i(0,0)|$
of  $|\phi_i(0,T)|$ ($i=1,2$) at successive reduced times $T (\equiv t/0.05)$, where
this error is
maximum, 
 calculated with $\Delta=0.05$ and $\Delta_1= 0.01$. The cases considered 
correspond to model (a) of Figs. 1 and 4
} \vskip .2cm

{\begin{center}{\begin{tabular}
{|c|c|c|c|c|c|c|c|c|c|c|c|c|c|c|c  |c |c |c |c|} 
\hline Fig. 1  &$T$ &0&19
 &  42 &67   & 101  & 125 & 165  &190
 & 225& 249&271& 311&337&371&395&414&458&483\\ $\phi_1$(a)
& ${\cal E}$ &0&$-1.7$  & 2.5 &$-1.0$  & 3.5 &$-1.4$
& 3.2 &$-1.9$
&$2.9$&$-2.7$ &2.1&$-3.2$&4.0&$-2.1$&$4.2$&$-1.8$&5.4&$-3.7$\\
\hline Fig. 1  &$T$ &0
 & 36  & 85  & 163  &219  &  275 &353
 & 395&477 & & && & && & &\\ $\phi_2$(a)
& ${\cal E}$ & 0 & 5.4 & $-4.9$ & 6.2  &$-4.2$
& $5.5$&$-4.2$
&$6.2$ &$-4.6$ & & && & && & &\\
\hline Fig. 4  &$T$ &0
 & 13  & 110  & 192  &250  &  318 &
376 &467 & & & & & & & & & &\\ $\phi_1$(a)
& ${\cal E}$ & 0 &1.2  & $-0.6$ & 1.0  &$-0.5$
&1.0 &$-0.3$
&0.9 & & & & & && & && \\
\hline Fig. 4  &$T$ &0
 & 27  & 113  & 183  & 245 &  323 &
386 & 462 & & & && & & & & &\\ $\phi_2$(a)
& ${\cal E}$ &0  & 3.1 & $-2.7$ & $3.2$ &$-2.5$
&$3.2$ &$-1.7$ 
& 3.1  & & & & &&  &  & & &\\
\hline   
\end{tabular}}\end{center}} \vskip .2cm

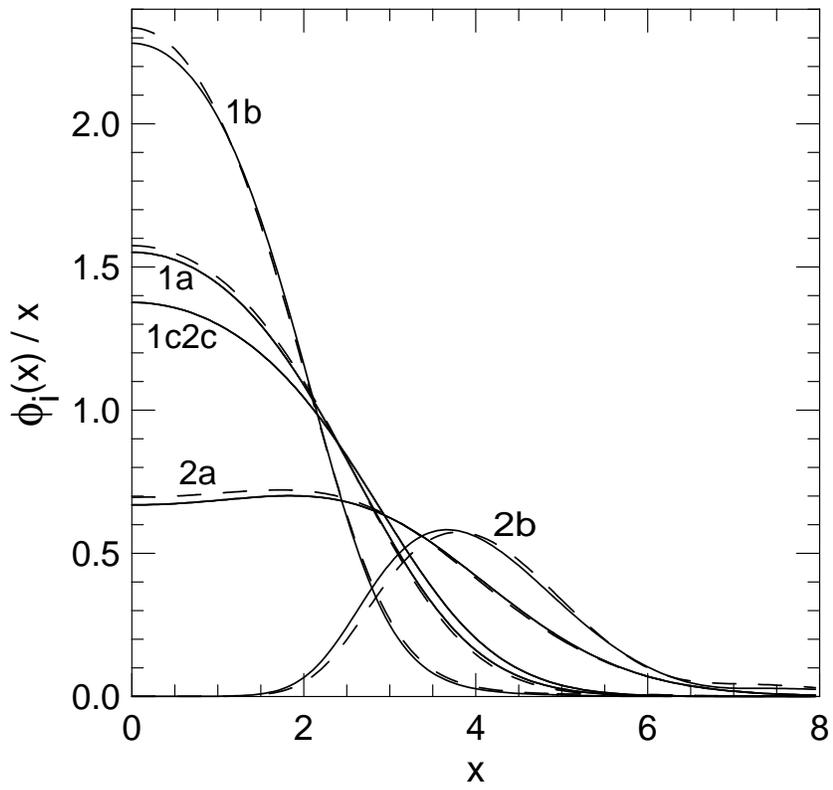
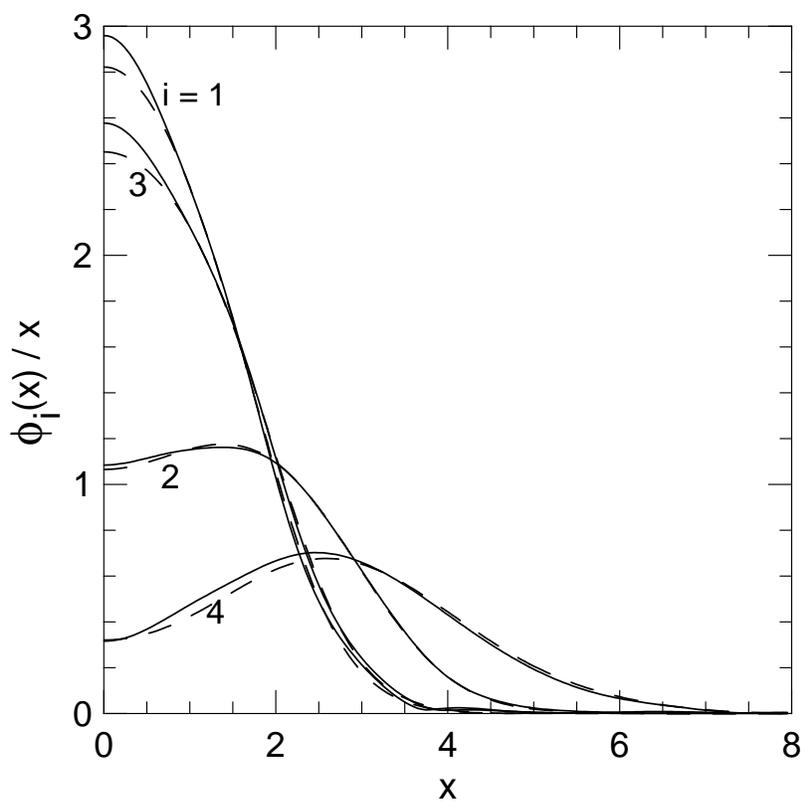
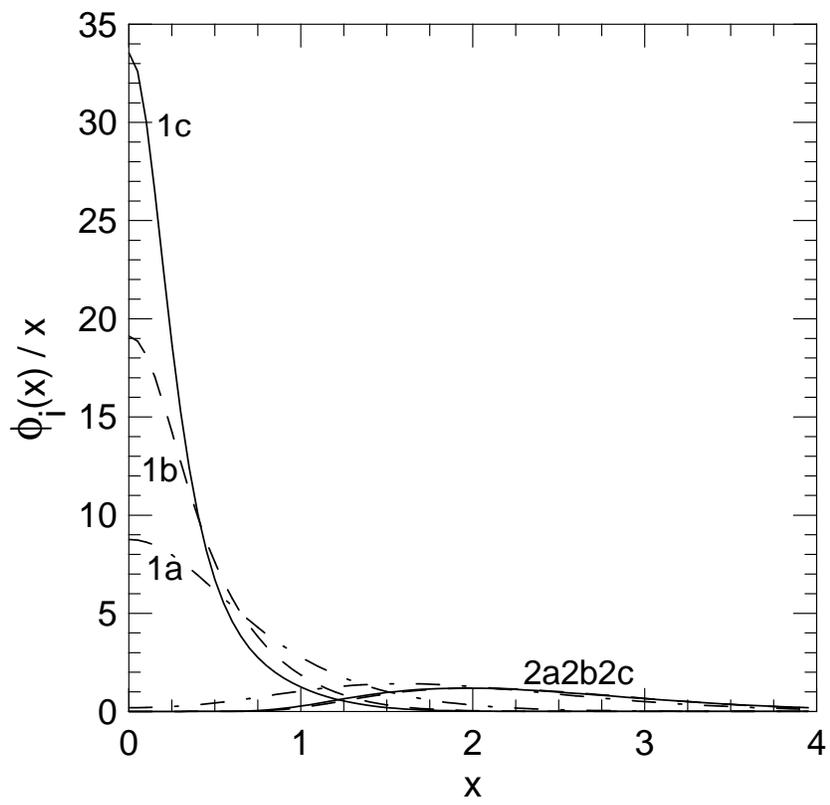
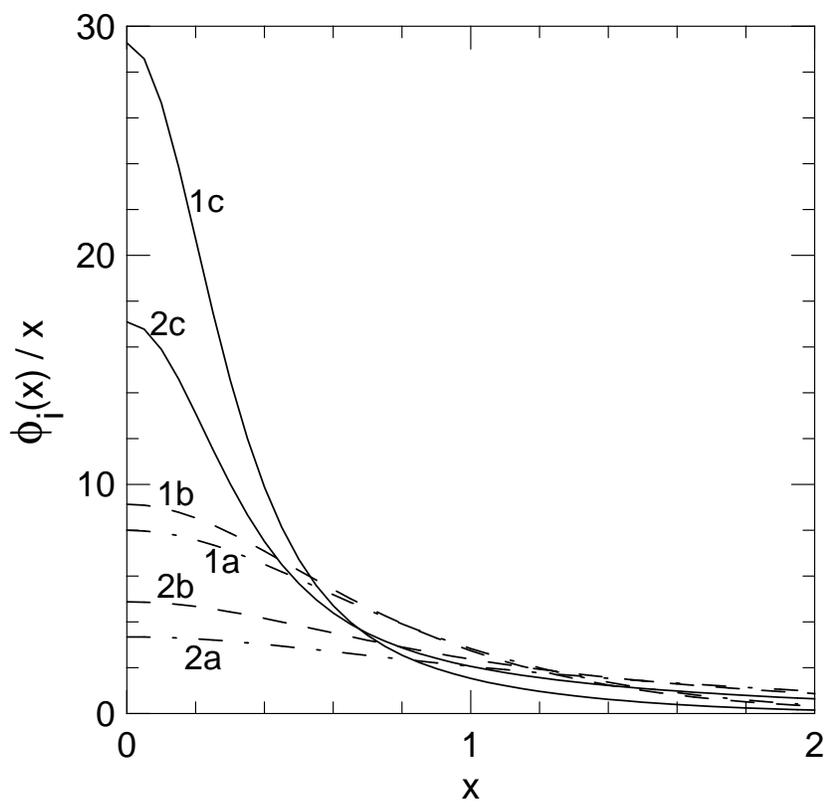
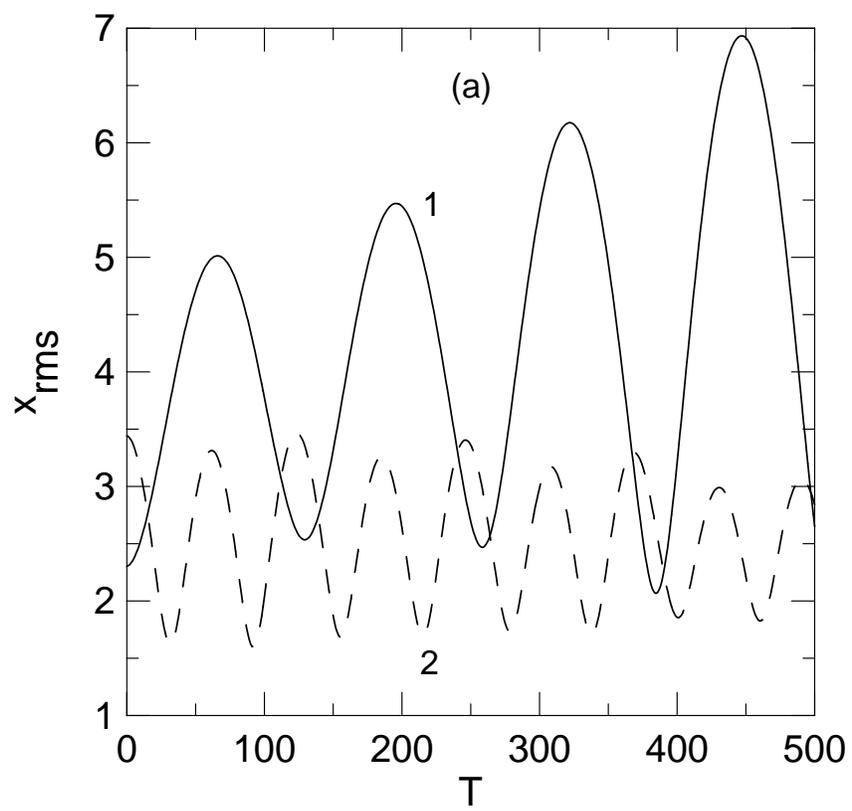
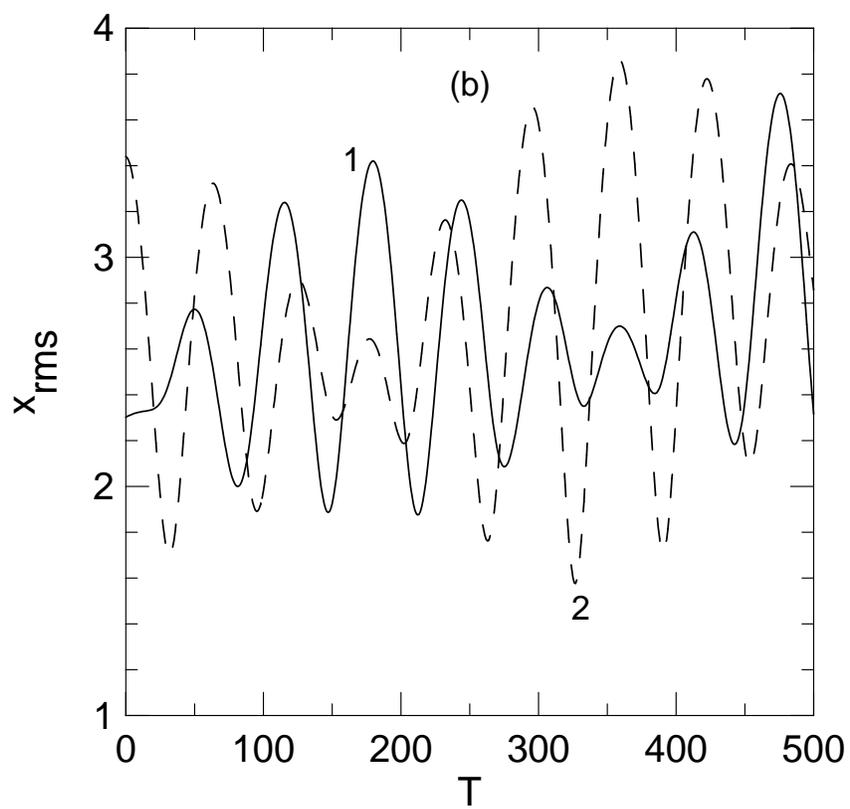
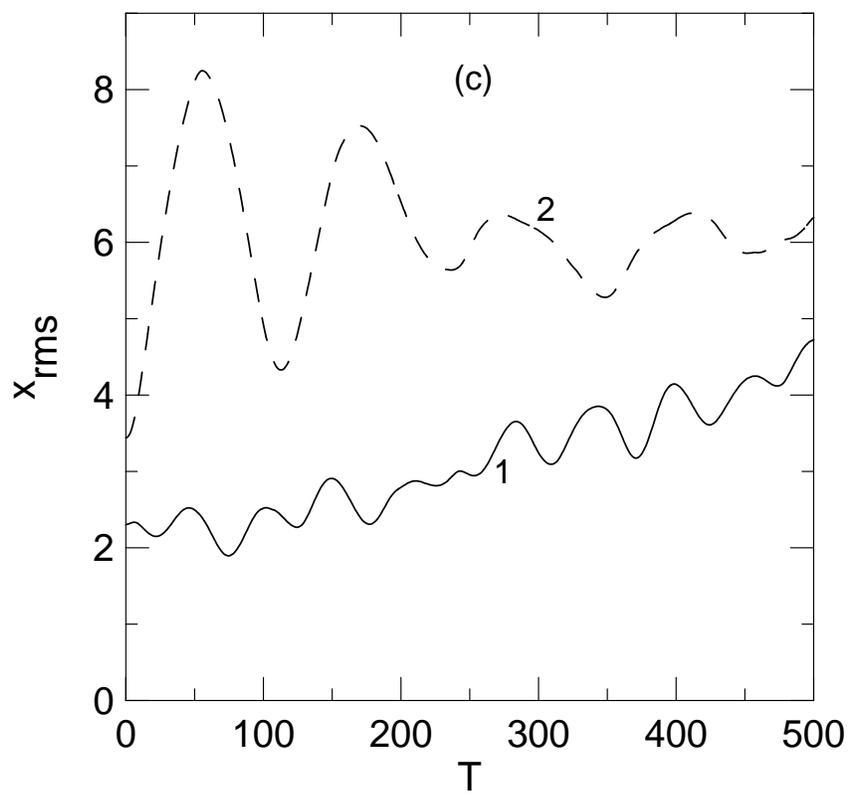
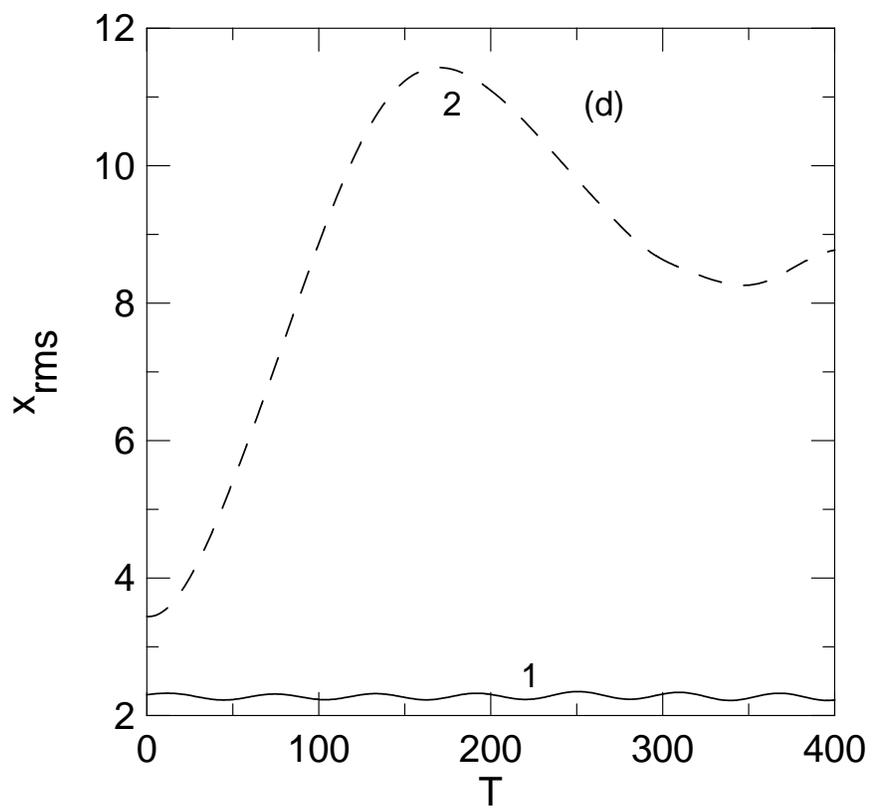
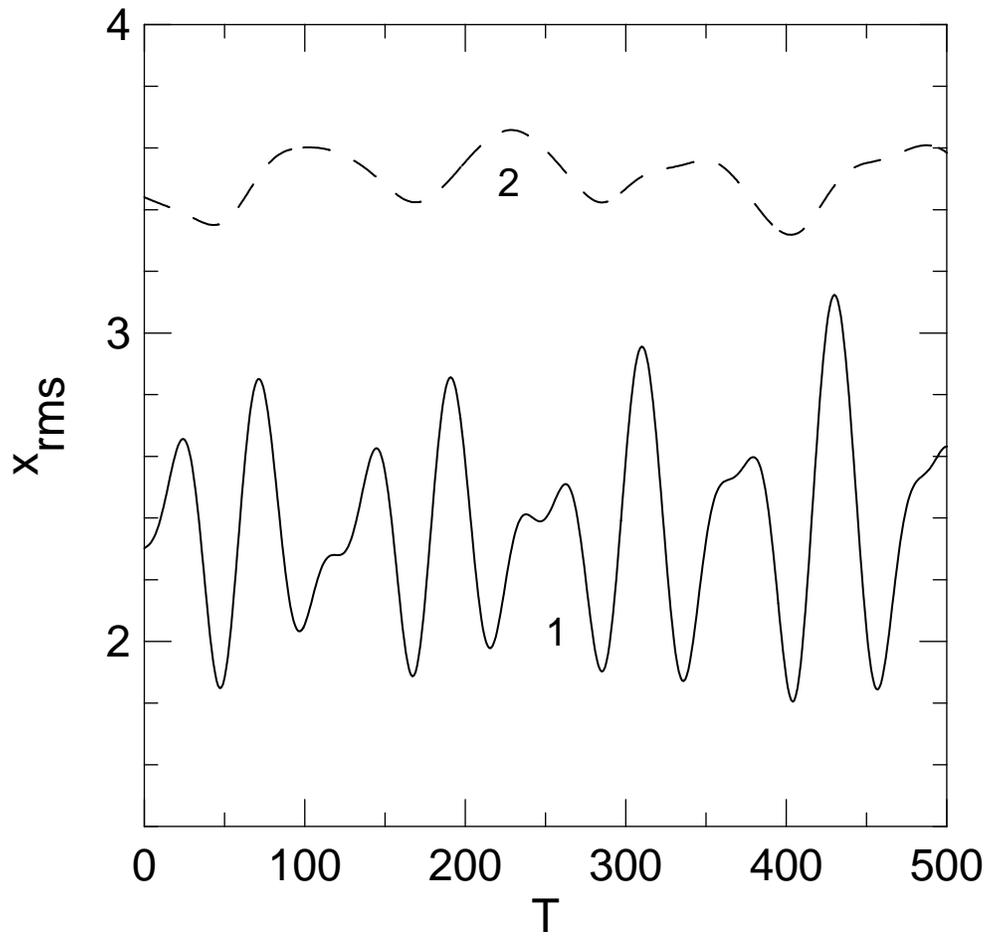
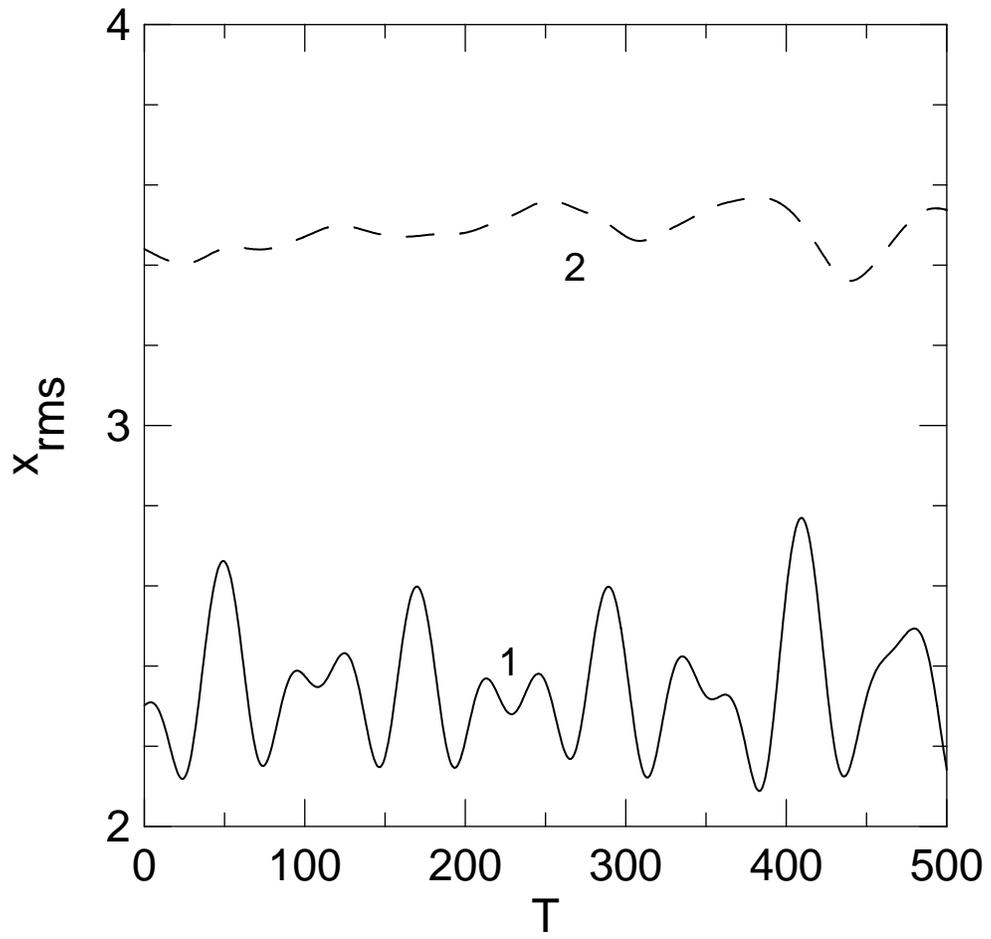
\end{document}